# Generation of nonparaxial self-accelerating beams using pendant droplets

Qiyue Zhang,[1,2] Peng Zhang,[1] Huizhong Xu,[1] Weining Man,[1] and Zhigang Chen[1,3,a]

**AFFILIATIONS**

[1]Department of Physics and Astronomy, San Francisco State University, San Francisco, CA 94132, USA
[2]Westview High School, Portland, OR 97229, USA
[3]School of Physics and TEDA Applied Physics Institute, Nankai University, Tianjin 300457, China USA

[a]**Author to whom correspondence should be addressed:** zhigang@sfsu.edu

**ABSTRACT**

We propose and demonstrate the effectual generation and control of nonparaxial self-accelerating beams by using UV-resin pendant droplets. We show that the geometrical shape of the hanging droplets formed as a result of the interplay between surface tension and gravity offers a natural curvature enabling the generation of nonparaxial self-accelerating beams. By simply adjusting the tilt angle of the surface where the droplets reside, a passing light beam is set to propagate along different curved trajectories, bending into large angles with non-diffracting features superior to a conventional Airy beam. Such self-accelerating beams are directly traced experimentally through the scattered light in yeast-cell suspensions, along with extensive ray tracing and numerical simulations. Furthermore, by modifying the shape of uncured pendant resin droplets in real time, we showcase the dynamical trajectory control of the self-accelerating beams. Our scheme and experimental method may be adopted for droplet-based shaping of other waves such as microfluidic jets and surface acoustic waves.

## 1. INTRODUCTION

The formation of liquid droplets is one of the most common phenomena in our daily life, from a dripping faucet to a spilled beverage on a dining table. Such a natural process contains rich physics, and has been a subject of intense interest for chemical engineering, drug delivery, and biomedical and microelectromechanical devices, to name just a few [1]. Optofluidics technology, which emerged as a fusion of microfluidics and optics, has been extensively studied in the past fifteen years [2-6]. For example, a variety of reconfigurable and responsive droplet-based optical devices for shaping and manipulating light have been proposed and demonstrated [3-6]. On the other hand, structured light [7,8], and particularly the non-diffracting self-accelerating beams [9-14], have been intensively studied over the years for their fundamental physics and many fascinating applications such as in remote spectroscopy [15], micro-particle manipulations [16-18], and biomedical imaging [19,20]. In practice, the generation of self-accelerating beams often involves expensive optical devices or sophisticated mask designs [10,12,21]. The caustic nature of the self-accelerating beams has also been employed for designing self-bending beams along desired and arbitrary trajectories [22,23]. In fact, natural caustic phenomena associated with droplets have intrigued scientists for many decades [24-26]. Liquid droplets have commonly been used to study the catastrophe theory behind the formation of various caustic patterns in transverse planes [27,28]. Recently, the development of elastomeric optics enabled by pendant (hanging) droplets has allowed for a convenient way to fabricate high-performance optical lenses [29,30]. As such, one may wonder: is it possible to use the droplet caustic phenomena to realize self-accelerating beams and dynamically control their trajectories?

In this work, we propose and demonstrate the generation of self-accelerating beams using tilted pendant droplets of UV resin. We show that, due to the interplay between gravity and surface tension, light can be focused and propagates along different curved trajectories after passing through droplets formed at different tilt angles. Such self-bending beams manifest the caustics in the longitudinal plane. By direct comparison with well-known Airy beams of a similar size, we demonstrate the nonparaxial nature of the self-bending beams created with the pendant droplets. Furthermore, dynamical control of their trajectories has also been successfully realized. Our method may find practical applications in a variety of fields, including optical manipulations and optofluidics, and in wave systems beyond optics.

## 2. METHOD AND EXPERIMENTAL SETUP

*2.1 Method*

Given the caustic nature of the self-accelerating beams [22,23], the basic idea for generating self-bending beams with droplets lies in properly harvesting the curvature of a pendant droplet formed under the influence of both gravity and surface tension. Our method is schematically shown in Fig. 1. Without any tilt, the pendant droplet serves as a high-quality lens as observed in previous experimental studies [29,30], which can focus light onto a single point [see Figs. 1(a) and 1(c)]. By gradually tilting the pendant droplet as shown in Fig. 1(b), light rays become realigned into a curved line caustic in the longitudinal plane, which represents a self-accelerating beam [see Fig. 1(d)].

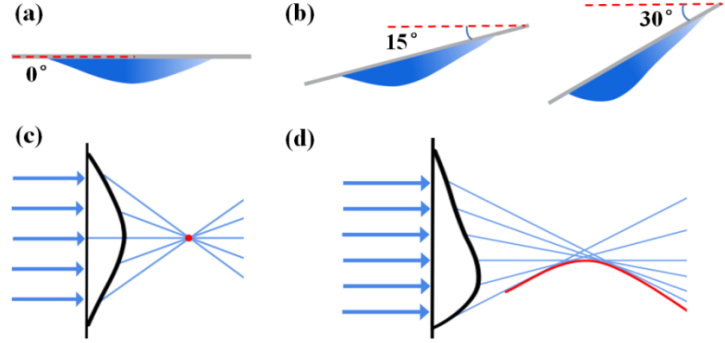

Fig. 1. Illustration of schemes for the generation of self-accelerating beams with pendant droplets. (a)-(b) Droplets under different tilt angles; (c) and (d) Ray diagrams for light passing through the droplets without and with a tilt angle, respectively.

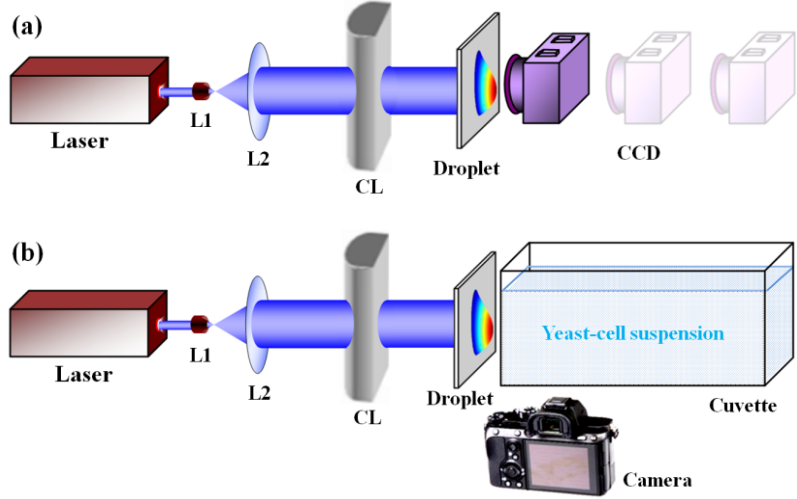

Fig. 2. Experimental setup for observation of output patterns of self-accelerating beams generated with tilted pendant droplets via a CCD (a), and for observation of side-view beam trajectories with a camera via imaging of the scattered light from a yeast-cell suspension (b).

*2.2 Experimental setup*

Our experimental setup is depicted in Figs. 2(a) and 2(b). Commercially available clear (transparent) UV resin is used, which has a refractive index of about 1.7. After a droplet of ~75μl is dispensed onto a microscope glass coverslip, it is immediately flipped and carefully mounted on a rotational stage to be cured under UV light. To experimentally observe the light dynamics passing through a droplet lens, we send an expanded Gaussian beam ($\lambda$ = 488 nm) through a cylindrical lens ($f$ = 200 mm) to form a light sheet that matches the central cross-section of the droplet. A CCD camera is placed right behind the droplet to measure its profile [see Fig. 2(a)]. To visualize the curved trajectories of the beam, we remove the CCD camera and place behind the droplet a 7.6-cm-long glass cuvette containing about 3-μm-sized 0.02 w.t.% yeast cells suspended in water [see Fig. 2(b)], and image the scattered light from the suspension with a camera.

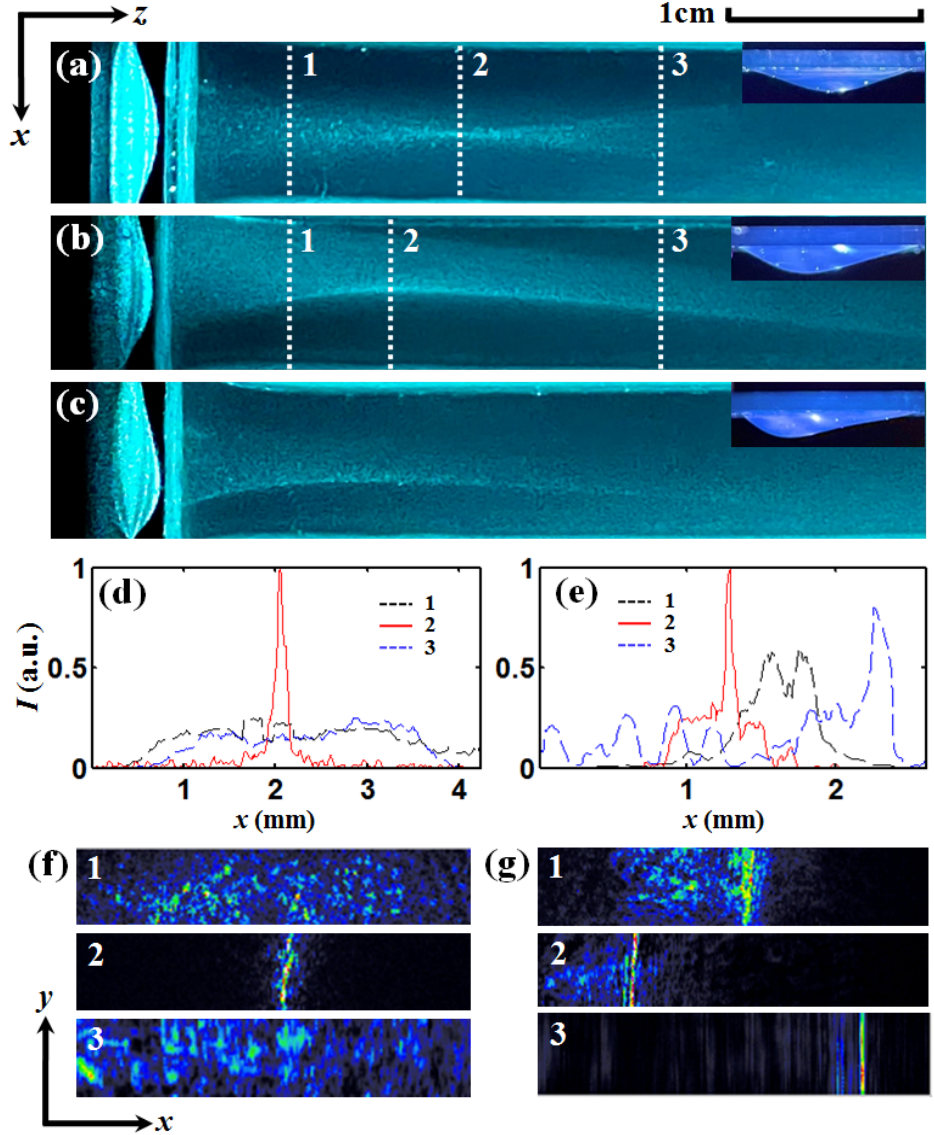

Fig. 3. Experimental side-views and transverse intensity profiles of a light beam after propagating through the UV-cured resin pendant droplets. Photographs in (a)-(c) are side-view images of the scattered light, and top-right insets are the pictures of the corresponding droplets formed at 0º, 15º, and 30º tilt angles. For comparison, (d) and (e) depict transverse intensity profiles recorded by the CCD camera, where the black, red, and blue curves show the intensity distribution at different z-positions marked by the white dashed lines 1, 2, and 3 in (a) and (b), respectively. Figures (f) and (g) are the corresponding output transverse intensity patterns taken at different propagation distances, showing clearly an Airy-type asymmetrical profile (g) produced by the asymmetric droplet (b) as opposed to that in (f) produced by the symmetric droplet (a).

## 3. EXPERIMENTAL RESULTS

Our experimental results are presented in Fig. 3, which highlight the trajectory and intensity pattern of a light beam after passing through a pendant droplet. Figures 3(a)-3(c) provide side-views of the light scattered in the yeast-cell suspension with the beam propagating along the z-direction. The insets show pictures of the cured droplet lenses. Figures 3(d) and 3(e) depict intensity distributions of the beam measured at different z-positions. We can clearly see that, when the pendant droplet is not tilted, it serves as a symmetric lens and focuses the light beam onto a single spot. In contrast, when the droplet is properly tilted, it turns into an asymmetric lens, and the path of the light beam is shaped into a curved trajectory. This is shown in Figs. 3(b) and 3(c), corresponding to a tilt angle of 15º and 30º, respectively. The transverse intensity patterns recorded by the CCD camera reveal more details of the beam structure. For the symmetric case, the beam profile is always centered– i.e. no self-bending is present, and the beam expands rapidly as it departs from the focal point. While for the tilted case, the beam size of the main lobe is well maintained even after propagating over several centimeters, representing diffraction-free behavior exhibited by a self-accelerating beam. More interestingly, the main lobe of the beam shifts several millimeters in the vertical direction, exhibiting a large-angle self-bending feature. Note that the fine structure in the "tail" of the generated beam is not revealed very clearly here in both side-view images and CCD recorded profiles due to our experimental limitations (such as poor camera and CCD resolutions), but the Airy-type asymmetrical profile is evident in Figs. 3(e) and 3(g), in contradistinction with the Gaussian-type propagation through a symmetric droplet. This point will be better manifested in the numerical simulations presented in the next section.

## 4. NUMERICAL SIMULATIONS

The equilibrium shapes of liquid droplets are typically obtained by numerically solving the Young-Laplace equation [31,32]. Very recently, a few fairly accurate analytical models describing the shapes of pendant droplets under different tilted conditions have been published [33-35]. In essence, the geometric shape of a pendant droplet in the spherical coordinate system can be approximated by the following equations [35]:

$$r(\theta,\varphi) = 1 + Br_{01}(\theta)\cos\alpha + Br_{11}(\theta)\sin\alpha\cos\varphi \tag{1a}$$

$$r_{01}(\theta) = \frac{\cos\theta - \cos\theta_0}{6} + \frac{\cos\theta}{3}\log\frac{1+\cos\theta_0}{1+\cos\theta} \tag{1b}$$

$$r_{11}(\theta) = \frac{\sin\theta}{3}\left[\frac{\cos\theta}{1+\cos\theta} - \frac{\cos\theta_0}{1+\cos\theta_0} + \log\frac{1+\cos\theta}{1+\cos\theta_0}\right] \tag{1c}$$

where $\alpha\in[0,\pi]$ is the tilt (slope) angle, $\theta\in[0, \theta_0]$ is the polar angle with $\theta_0$ being the contact angle, $\varphi\in[-\pi, \pi]$ is the azimuth angle, and $B$ is the Bond number measuring the strength of gravity relative to the surface tension.

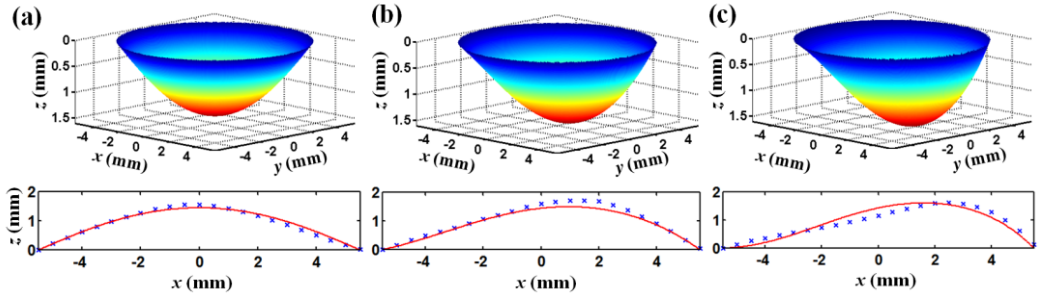

Fig. 4. Best-fit analytical solutions of the shapes of the pendant droplets for the cured droplets at 0º, 15º, and 30º tilt angles, corresponding to the cases shown in Figs. 3(a)-3(c). Top: Three-dimensional display of the droplet shape; Bottom: Data points tracing the edges of the droplet images (blue crosses) vs. analytical simulation results of the droplet profiles across the droplet center (red solid curves).

To perform numerical simulations, we first trace out the three geometry shapes of the droplets as shown in the insets of Figs. 3(a)-3(c). Then, we numerically fit the data points of these droplet shapes with the theoretical model described by Eq. (1). The volume difference among the three droplets is negligible, hence for simplicity, we assume they have the same contour during initial contact with the glass coverslip. As a result, the same $B$ and $\theta_0$ values will be used for all three droplets in our simulations. The best-fit parameters we have obtained are $B$=14.36 and $\theta_0$=27.5º. Detailed results are shown in Fig. 4, where Figs. 4(a), 4(b), and 4(c) correspond to a tilt angle of 0º, 15º, and 30º, respectively. The simulated three-dimensional droplet shapes are displayed in the top panels, and the bottom panels depict the data points (blue crosses) tracing the droplet images and the analytical profiles (red solid curves) across the center of the droplets. The deviation of the experimental data points from the analytical shapes is mainly attributed to imperfect droplet preparation and inaccuracy during the imaging processes.

By utilizing Eq. (1) with the best-fit parameters, we perform ray tracing and BPM (Beam Propagation Method) simulations of light propagation after passing through the droplets. The results are shown in Fig. 5, where Figs. 5(a)-5(c) display the ray tracing and BPM (insets) results for the cases at 0º, 15º, and 30º tilt angles, respectively. Note that the BPM results are at the same scale as the ray tracing results, but are shifted along the positive x-axis for displaying purposes. The intensity distributions along the dashed lines in Figs. 5(a) and 5(b) are plotted in the top and bottom panels of Fig. 5(d), respectively. Again, the intensity profiles shown in the bottom panel of Fig. 5(d) contain artificial shifts along the *x*-axis for displaying purposes. From Fig. 5(a), we can clearly see that without any tilt, all the light rays are focused into a single point. In the presence of a tilt angle, however, the light rays are smoothly redistributed as tangents of a curved trajectory. Such an intriguing behavior is caused by the natural curvature of the droplet surface due to the interplay between surface tension and gravity, as manifested by the Bond number $B$ in our model. In addition, the perfect focusing case at a 0º tilt angle represents a highly optimized yet unstable state, since any deformation introduced to the symmetric shape of the droplets would readily cause the caustics to reform. In addition, the intensity profiles shown in Fig. 5(d) clearly unveil the fine structure of the beam along its tail, indicating the non-diffracting feature of the self-bending beam. Overall, the numerical simulations match well with our experimental observations presented in Fig. 3.

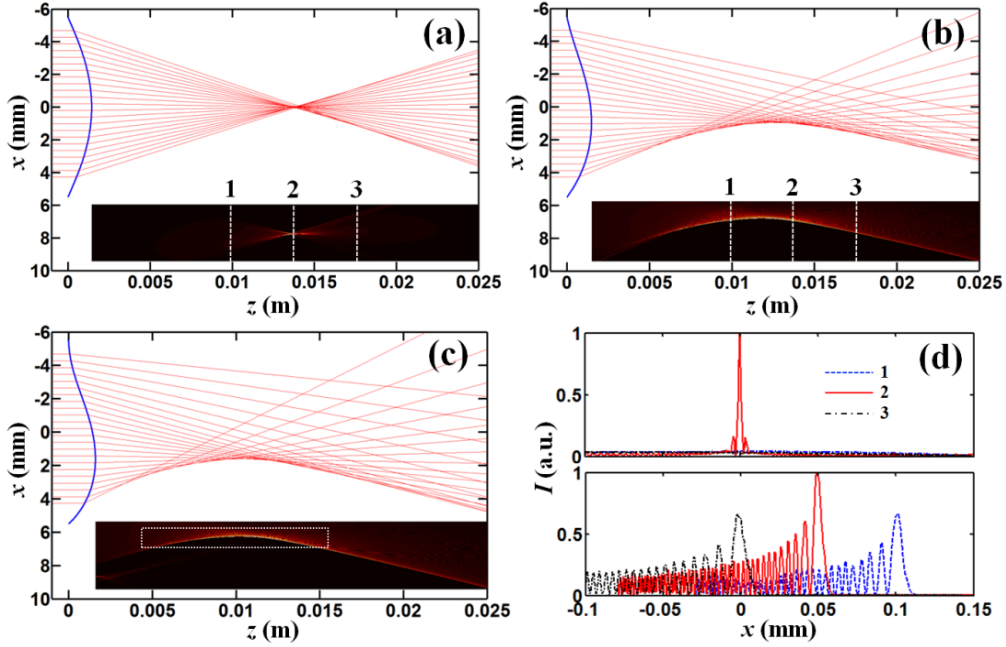

Fig. 5. (a)-(c) Ray tracing and BPM simulations corresponding to the cases shown in Fig. 2. (a)-(c) correspond to the propagation dynamics of a beam exiting the cured droplets at 0º, 15º, and 30º tilt angles, respectively. Note that the BPM simulation results have been artificially shifted along the positive *x*-axis for displaying purposes. (d) Intensity profiles along the white dashed lines in (a) and (b), where the top and bottom panels correspond to (a) and (b), respectively. Again artificially shifts are applied along the *x*-axis in the bottom plot of (d).

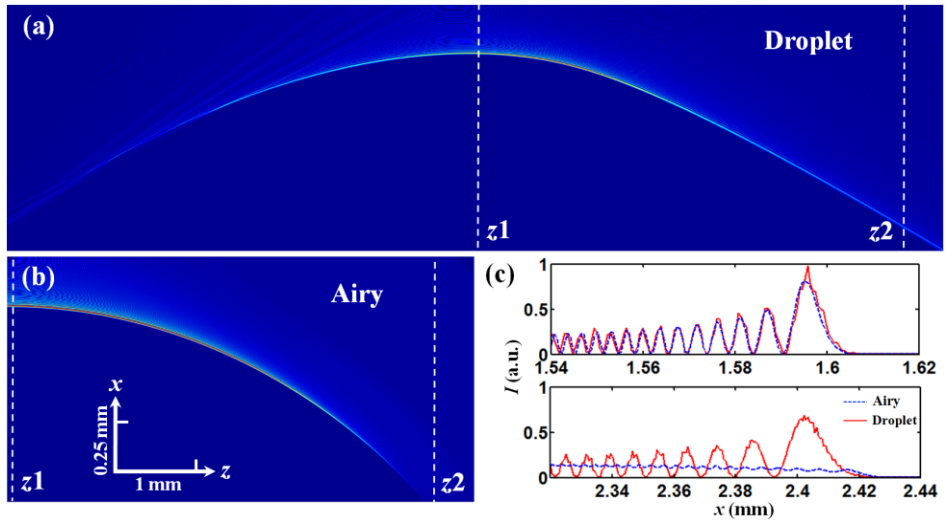

Fig. 6. Direct comparison between the self-bending beam generated from the droplet as shown in Fig. 5(c) and a conventional Airy beam with a similar main-lobe size. (a) and (b) are side-view propagations of the two beams, where (a) is a close-up view of the beam structure within the white dashed rectangle marked in Fig. 5(c). Figure (c) displays the vertical intensity profiles along the dashed lines in (a) and (b), where the top and bottom panels correspond to the left (at $z1$) and right (at $z2$) white dashed lines, respectively.

To further demonstrate the nonparaxial nature of the self-accelerating beams generated with the droplets, we compare the beam propagation dynamics shown in Fig. 5(c) with an Airy beam of a similar main-lobe size. The results are depicted in Fig. 6, where Fig. 6(a) shows a close-up view of the area marked by the white dashed rectangle in Fig. 5(c), and Fig. 6(b) displays the side-view of a conventional Airy beam propagation [10-14]. The intensity profiles along the dashed lines marked at $z1$ and $z2$ in Figs. 6(a) and 6(b) are superimposed in Fig. 6(c) for a direct comparison, where the top and bottom panels correspond to the left ($z1$) and right ($z2$) white dashed lines, respectively. It is evident from Fig. 6(c) that the non-diffracting property of the Airy beam breaks down at large bending angles due to its intrinsic paraxial limitation, however, the self-bending beam exiting the pendant droplet maintains well its main lobe even when propagating into the nonparaxial regime.

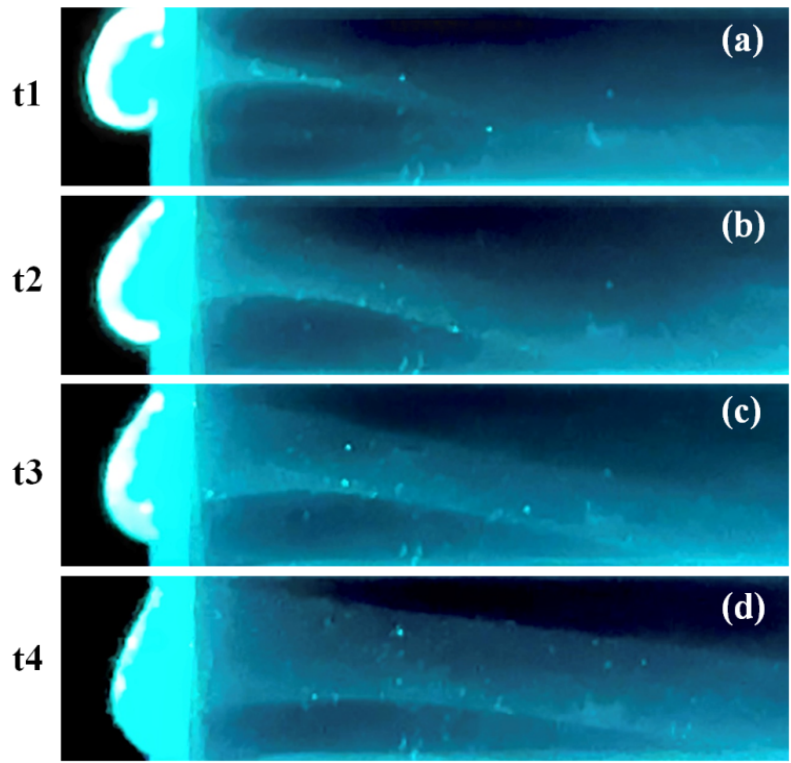

Fig. 7. Dynamical control of the beam trajectory via morphing pendant droplets under the action of gravity. (a)-(d) are the side-view snapshots taken from a recorded video at different moments as an uncured UV-resin droplet slides down the wall of a cuvette.

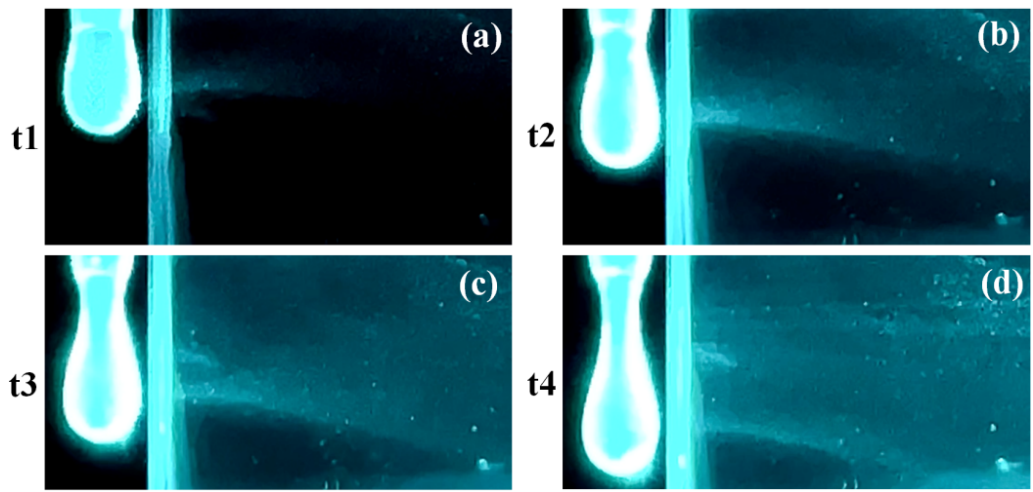

Fig. 8. Active control of the beam trajectory via morphing pendant droplets under the action of controlled airflow. (a)-(d) Snapshots taken from a recorded video at different moments, showing side-view propagations of a light beam passing through an uncured UV-resin droplet forming slowly at the opening of a capillary tube as it is being blown with controlled airflow.

## 5. DYNAMICAL TRAJECTORY CONTROL

Given the natural connection between our method and those used in optofluidics [2-6], the capability of dynamically manipulating light with droplets is obvious. Next, we demonstrate that the curved beam trajectory can be dynamically changed with uncured UV resin pendant droplets. One example is shown in Figs. 7(a)-7(d), which are snapshots taken from a recorded video at different moments when a droplet falls on the side wall of a cuvette. Figure 7(a) shows the early stage when the beam starts to transition from a single-point focusing case to a self-accelerating case. Figures 7(b)-7(d) show that both the curved angle and range of the trajectory can be tuned via morphing the droplet under the action of gravity. Another method, more active and controllable, is illustrated in Figs. 8(a)-8(d), where the resin is injected into a capillary tube so the droplet shape can be modified in real-time via controlled airflow. It should be pointed out that the experimental results presented here merely serve as a proof of principle, yet they can certainly be optimized in better-controlled experiments with optofluidic techniques.

## 6. CONCLUSIONS

To summarize, we have proposed and demonstrated a simple yet effective scheme for the generation of large-angle self-accelerating beams with tilted pendant droplets. By simply using tilted UV-resin pendant droplets, a light beam can be set to propagate along different curved trajectories in a non-diffracting fashion. Such large-angle self-bending beams have been directly observed through the scattering of light in a yeast-cell suspension. We have shown that such generated self-accelerating beams can outperform the conventional Airy beams due to their nonparaxial nature. By modifying the shape of uncured pendant resin droplets in real time, we have experimentally demonstrated the dynamical control of the trajectories of self-accelerating beams. The propagation dynamics and caustic nature of such beams are examined by ray tracing based on the geometrical shape of the hanging droplets, and experimental results of side-view beam propagation are well corroborated by the BPM simulations. Our method paves the way for droplet-based beam shaping, which may find many practical applications in optical trapping, particle manipulations, and optofluidics.

**ACKNOWLEDGEMENTS**

We thank Elodie Naimi for her assistance.

**DISCLOSURES**

The authors declare no conflicts of interest.

**DATA AVAILABILITY**

Data that support the findings of this study are available within the article.